\documentstyle[12pt]{article}

\begin{document}
\vspace*{2cm}
\noindent{\large\bf
Diffusion-controlled reactions in presence of polymers}
\vspace{1.5ex}

\noindent
Ch. von Ferber$^a$ and Yu. Holovatch$^b$
\vspace{1.5ex}

\noindent
$^a$
Institut f\"ur Theoretische Physik II, Heinrich-Heine-Universit\"at
D\"usseldorf, \\
D-40225 D\"usseldorf, Germany

\vspace{1.5ex}

\noindent
$^b$
Institute for Condensed Matter Physics, Ukrainian Acad. Sci.,\\
1 Svientsitskii Str., UA-290011, Lviv, Ukraine 
\vspace{2ex}

We study the properties of diffusion controlled reactions with traps
or reaction sites attached to polymer chains or to a star 
polymer. Using a field theoretical renormalization group approach we
obtain the scaling of the moments of concentration of diffusing particles 
near the core of a star polymer
and calculate numerical values of the exponents governing 
the scaling of the reaction rate. 

\vspace{2ex}

Chemical reactions between diffusing reactants play an important role. 
Examples of such reactions can be found in different systems, ranging from 
biological systems to nuclear reactors (see e.g. \cite {Rice85} and 
references therein). One more place where these reactions appear as
a limiting stage are aggregation models \cite{agr}. Of particular interest
are reactions between reactants of different nature: particles $A$ which 
freely diffuse in a solution and particles $B$ which are attached to 
polymer chains immersed in the same solvent \cite{Oshanin} (the concentration
of polymers being low enough to allow to neglect the inter-chain interaction).
Such a process may be considered also as a trapping reaction of $n$ 
particles of $A$ type and  traps $B$:
\begin{equation}
\label{1}
A^n + B \rightarrow 0.
\end{equation}
The reaction rate $k_n$ of (\ref{1}) 
in the vicinity of a certain trap on the polymer of size $R$ 
is proportional to the averaged moments 
of the concentration $\rho$ of diffusing particles near this trap. 
It scales with $R$ as \cite{Cates}: 
\begin{equation}
\label{2}
k_n \sim <\rho^n> \sim (R/l)^{-\lambda_n},
\end{equation}
with $l$ being a characteristic length scale.

The goal of this paper is to investigate the scaling properties (\ref{2})
in the vicinity of an absorber in form of a star-shaped polymer.
We are interested in the scaling properties of (\ref{2}) near the trap 
placed at the center of a polymer star, provided all polymer chains are
decorated by traps and particles meeting them are absorbed.
As a model we consider the polymer chains to be either random walks (RWs)
or self -avoiding walks (SAWs).
To this end we generalize \cite{FerHol} the idea of Cates and Witten 
\cite{Cates} to  describe both the polymer absorber and the random 
walks of diffusing particles in the same way. 
In a steady state limit the diffusion equation describing particles
$A$ is reduced to the Laplace equation. 
In terms of the path integral solution of the Laplace equation one
finds that $\rho(r)$ at point $r$ near the absorber is proportional to the
number of RWs that end at point $r$ and avoid the absorber
\cite{Cates,FerHol}.
The $n$th power of $\rho(r)$ is proportional to the $n$th power of
the above mentioned number, i.e. it is defined by a partition function
of a star of $n$ walks \cite{stars}.
Furthermore, introducing the mutual avoidance
conditions between the `$n$-walk star' and the $m$-polymer star (representing
the absorber) one has to calculate the partition function of a co-polymer
star consisting of chains of two different species that avoid each other.
These correspond to the
trajectories of diffusing particles (being RW) and the absorbing polymer
which for the purpose of present study is chosen as a RW or a SAW.
Making use of the previously developed \cite{FerHol}
theory of copolymer stars and networks and mapping the model of
co-polymer stars to the appropriate field theory
\cite{ternary,FerHol}
one may relate the spectrum of exponents (\ref{2})
to the exponents that define the  scaling properties of co-polymer
stars. In particular, for a co-polymer star consisting of $m$ chains
of one species and $n$ chains of the other species the number of
configurations $Z_*$ scales as \cite{FerHol}:
\begin{equation} \label{3}
Z_* \sim (R/l)^{\eta_{mn}-m\eta_{20} -n\eta_{02}}.
\end{equation}
Here, the $\eta_{mn}$ represent the co-polymer star
exponents. The latter have been calculated using a field theoretical
renormalization group approach
\cite{Zinn89} and are known in the third order of
perturbation theory \cite{FerHol}.
By means of a
short-chain expansion \cite{Ferber97} the set of exponents
$\eta_{mn}$ (\ref{3}) can be related to the exponents $\lambda_{mn}$
(\ref{2}), that govern the scaling of the rate of reaction (\ref{1})
in the case of an absorber in the form of an $m$-arm star.
Considering the absorber to be
either a RW star or a SAW star we define the exponents:
\begin{eqnarray} \label{4}
\lambda^{RW}_{mn} =-\eta^{G}_{mn}, \\
\nonumber
\lambda^{SAW}_{mn} =
-\eta^{U}_{mn}+\eta^{U}_{m0},
\end{eqnarray}
here, we keep notations of
\cite{FerHol}
for the expressions for
exponents $\eta$ defined in different fixed points of the
renormalization group transformation. 
\begin{table}[h]
\caption{ \label{tab1}
Values of the exponent $\lambda^{RW}_{mn}$
at $d=3$ obtained by $\varepsilon$-expansion
($\varepsilon$) and by fixed dimension technique
($3d$).
}
\tabcolsep1.4mm
\begin{tabular}{lrrrrrrrrrrrr}
\hline
$m$ &
\multicolumn{2}{c}{$1$}&
\multicolumn{2}{c}{$2$}&
\multicolumn{2}{c}{$3$}&
\multicolumn{2}{c}{$4$}&
\multicolumn{2}{c}{$5$}&
\multicolumn{2}{c}{$6$}\\
$n$ &
$\varepsilon$ & $3d$ &
$\varepsilon$ & $3d$ &
$\varepsilon$ & $3d$ &
$\varepsilon$ & $3d$ &
$\varepsilon$ & $3d$ &
$\varepsilon$ & $3d$ \\
\hline
1 &
  0.56 &  0.58 &
  1.00 &  1.00 &
  1.33 &  1.35 &
  1.63 &  1.69 &
  1.88 &  1.98 &
  2.10 &  2.24 \\
2 & & &
  1.77 &  1.81 &
  2.45 &  2.53 &
  3.01 &  3.17 &
  3.51 &  3.75 &
  3.95 &  4.28 \\
3 & & & & &
 3.38 &  3.57 &
 4.21 &  4.50 &
 4.94 &  5.36 &
 5.62 &  6.15 \\
4 & & & & & & &
 5.27 &  5.71 &
 6.24 &  6.84 &
 7.12 &  7.90 \\
5 & & & & & & & & &
 7.42 &  8.24 &
 8.50 &  9.54 \\
6 & & & & & & & & & & &
 9.78 &  11.07\\
\hline
\end{tabular}
\end{table}
Based on these expressions we
get the result for the exponents in the form of an $\varepsilon$-expansion
series:
\begin{eqnarray}
\lambda^{RW}_{mn} (\varepsilon) &=& m n  \frac{\varepsilon}{2} +
m  n  (3-m-n)\frac {\varepsilon^{2}}{8}[ 1 - 
(m+n+3\zeta(3)-3)\frac {\varepsilon}{2}] ,
\label{5}
\end{eqnarray}
\begin{eqnarray} \nonumber
\lambda_{mn}^{SAW}(\varepsilon)&=&
3\,mn\frac{\varepsilon}{8}
-mn\left (18\,n+42\,m-91\right )\frac{\varepsilon^{2}}
{256} +mn[492\,{m}^{2}-2290\,m+2463+
\\
\label{6} &&
108\,{n}^{2}-1050\,n+540\,n\zeta (3
)+1188\,m\zeta (3)+504\,mn-2652\,\zeta (3)]\frac{{\varepsilon}^{3}}
{4096}.
\end{eqnarray}
Here, $\varepsilon=4-d$,  $d$ is the space dimension, and $\zeta(3)
\simeq 1.202$ is the Riemann zeta function.
Similarly, we obtain the expansions for (\ref{4}) in form of a 
series at fixed $d=3$
dimensions. Series of the type (\ref{5}),(\ref{6}) 
are known to be asymptotic at best \cite{Zinn89}. 
We obtain numerical values of the exponents $\lambda_{mn}$
applying Borel transformation improved by a conformal mapping procedure
\cite{Zinn89}. 
Tables \ref{tab1}  (\ref{tab2}) contain the results for the exponents
$\lambda^{RW}_{mn}$ ($\lambda^{SAW}_{mn}$) describing the scaling of the 
rate (\ref{2}) of the reaction  (\ref{1}) in the vicinity of a trap attached 
at the core of a polymer star made of $m$ RWs or SAWs 
respectively. The difference between the numbers obtained by resummed 
$\varepsilon$ and fixed $d$ -technique may also serve to test the 
accuracy of the data obtained. One can see the decrease of accuracy for
high functionalities of the star arms caused by the growth of the 
coefficients of the asymptotic 
expansions (\ref{5}),(\ref{6}) with $m,n$.
Let us note, that the case $m=2$ corresponds to a trap 
located on the chain polymer, whereas $m=1$ corresponds to a trap 
attached at the polymer extremity.  

\begin{table}[h]
\caption{ \label{tab2}
Values of the exponent $\lambda^{SAW}_{mn}$
at $d=3$ obtained by $\varepsilon$-expansion
($\varepsilon$) and by fixed dimension technique
($3d$).
}
\tabcolsep1.4mm
\begin{tabular}{lrrrrrrrrrrrrr}
\hline
$m$ &
\multicolumn{2}{c}{$1$}&
\multicolumn{2}{c}{$2$}&
\multicolumn{2}{c}{$3$}&
\multicolumn{2}{c}{$4$}&
\multicolumn{2}{c}{$5$}&
\multicolumn{2}{c}{$6$}\\
$n$ &
$\varepsilon$ & $3d$ &
$\varepsilon$ & $3d$ &
$\varepsilon$ & $3d$ &
$\varepsilon$ & $3d$ &
$\varepsilon$ & $3d$ &
$\varepsilon$ & $3d$ \\
\hline
1 & .43 &.45 &.70 &.70 &.89 &.91& 1.03& 1.09& 1.14& 1.24& 1.23& 1.39
\\ 
2 & .79 &.81& 1.30& 1.32& 1.69& 1.76& 1.97& 2.12& 2.21& 2.43& 2.41& 2.72 
\\
3 &1.09& 1.09& 1.85& 1.91& 2.41& 2.54& 2.84& 3.10& 3.21& 3.57& 3.53& 4.02
\\ 
4 &1.35& 1.37& 2.33& 2.43& 3.07& 3.28& 3.66& 4.02& 4.17& 4.67& 4.60& 5.27
\\ 
5 &1.60& 1.64& 2.77& 2.93& 3.69& 3.99& 4.44& 4.92& 5.08& 5.75& 5.63& 6.49
\\
6 &1.81& 1.89& 3.18& 3.40& 4.26& 4.66& 5.17& 5.77& 5.95& 6.78& 6.62& 7.68 
\\
\hline
\end{tabular}
\end{table}

Numerical values (tables \ref{tab1},  \ref{tab2})
and analytic expressions (\ref{5}), (\ref{6}) for the exponents governing the 
reaction rate in diffusion-controlled reactions involving star polymers 
are the main result of the present study. To conclude, let us analyze several 
particular cases:
\begin{itemize}
\item 
{\em For a given $m$-star absorber} of size $R$ the reaction rate 
(\ref{2}) scales as $k_{mn} \sim (R/l)^{-\lambda_{mn}}$. Increase of 
the size $R$ by a factor of $a$ results in  
$k^{\prime}_{mn} \sim (aR/l)^{-\lambda_{mn}}$
leading to:
\begin{equation}
\label{7}
k^{\prime}_{mn}/k_{mn} \sim a^{-\lambda_{mn}}
\end{equation}
with $\lambda_{mn}$ positive: increasing  $R$ by a factor of $a$ 
the reaction rate decreases $a^{-\lambda_{mn}}$ times. 
\item
{\em For a given reaction type } (\ref{1}) (i.e. for a fixed number $n$ of 
particles which are trapped simultaneously) attaching $m_1$ additional arms 
to an $m$-arm star absorber results in a decrease of the reaction rate:
\begin{equation}
\label{8}
k_{(m+m_1),n}/k_{mn} \sim (R/l)^{-(\lambda_{(m+m_1),n} -\lambda_{mn})},
\end{equation}
as far as $\lambda_{m_2n} > \lambda_{m_1n}$ for $m_2>m_1$.
\item
{\em For a given $m$-star absorber} the change of the type of reaction 
(\ref{1}) to $A^{n_1} + B \rightarrow 0, \, n_1>n$ results in a decrease 
of the reaction rate:
\begin{equation}
\label{9}
k_{m,(n+n_1)}/k_{mn} \sim (R/l)^{-(\lambda_{m,(n+n_1)} -\lambda_{mn})},
\end{equation}
as far as 
$\lambda_{mn_2} > \lambda_{mn_1}$ for $n_2>n_1$.
\end{itemize}

\vspace*{2ex}

We acknowledge helpful discussions with  L.~Sch\"afer. 
C.v.F. thanks the Deutsche Forschungsgemeinschaft for support
within SFB 237. Yu.H. gratefully acknowledges support by 
Deutsche Akademische Austauschdienst.

\end{document}